\documentclass[sigconf]{acmart}
\AtBeginDocument{%
  }

\usepackage{booktabs}
\usepackage{multirow}

\usepackage{listings}
\usepackage[most]{tcolorbox}
\usepackage{fancyvrb}
\renewcommand\footnotetextcopyrightpermission[1]{}

\begin{document}

\title{From Understanding to Action: Feedback-Grounded Policy Discovery for Generative Recommendation}



\author{Zhi Chen}
\authornote{This work was done during an internship at Kuaishou Technology.}
\authornote{Equal contribution.}
\affiliation{%
  \institution{Huazhong Agricultural University}
  \city{Wuhan}
  \country{China}
}
\email{chenzhi321654@gmail.com} 

\author{Minmao Wang}
\authornotemark[2]
\affiliation{%
  \institution{Fudan University}
  \city{Shanghai}
  \country{China}
}
\email{mmwang25@m.fudan.edu.cn}

\author{Xingchen Liu}
\affiliation{%
  \institution{Kuaishou Technology}
  \city{Beijing}
  \country{China}
}
\email{liuxingchen07@kuaishou.com}

\author{Haoqiang Liang}
\affiliation{%
  \institution{Kuaishou Technology}
  \city{Beijing}
  \country{China}
}
\email{lianghaoqiang@kuaishou.com} 

\author{Huihuang Lin}
\affiliation{%
  \institution{Kuaishou Technology}
  \city{Beijing}
  \country{China}
}
\email{linhuihuang@kuaishou.com} 

\author{Likang Wu}
\authornote{Corresponding authors.}
\affiliation{%
  \institution{Tianjin University}
  \city{Tianjin}
  \country{China}
}
\email{wulk@tju.edu.cn}

\author{Hongke Zhao}
\affiliation{%
  \institution{Tianjin University}
  \city{Tianjin}
  \country{China}
}
\email{hongke@tju.edu.cn}

\author{Yulong Wang}
\affiliation{%
  \institution{Huazhong Agricultural University}
  \city{Wuhan}
  \country{China}
}
\email{wangyulong6251@gmail.com}

\author{Shijie Yi}
\authornotemark[3]
\affiliation{%
  \institution{Kuaishou Technology}
  \city{Beijing}
  \country{China}
}
\email{yishijie@kuaishou.com}

\author{Fei Pan}
\affiliation{%
  \institution{Kuaishou Technology}
  \city{Beijing}
  \country{China}
}
\email{panfei05@kuaishou.com}

\author{Peng Jiang}
\affiliation{%
  \institution{Kuaishou Technology}
  \city{Beijing}
  \country{China}
}
\email{jiangpeng11@kuaishou.com}

\renewcommand{\shortauthors}{Chen et al.}
\begin{abstract}
Semantic-ID-based generative recommenders enable efficient next-item generation, but their item-level supervision mainly captures behavioral co-occurrence and local transitions. Large language models (LLMs) can complement these models by reasoning over heterogeneous interaction histories to understand the user's current demand. However, LLMs are not inherently trained with recommendation-specific outcome feedback, and linguistically plausible reasoning therefore does not necessarily lead to effective recommendation decisions. We term this mismatch the \emph{Understanding--Action Gap}. Accordingly, we distinguish \emph{intent knowledge}, which captures the user's current demand, from \emph{policy knowledge}, which specifies the recommendation direction and rejection boundary under that demand. To bridge this gap, we propose a feedback-driven agent framework that first induces task-oriented intent and then discovers recommendation policies according to their incremental utility over an intent-only baseline. Candidate policies are evaluated and refined using outcome-derived feedback rather than linguistic plausibility. We further transfer the resulting intent and policy knowledge into two latent tokens of a lightweight Semantic-ID generator through dual-space relational distillation, enabling LLM-free online inference. Experiments on public benchmarks show consistent improvements over baselines, while large-scale online A/B tests achieve gains of 4.506\% in Revenue and 4.621\% in ADVV.
\end{abstract}



\maketitle

\section{Introduction}

Generative recommendation~\cite{zhai2024actions,ETEGRec,llm4rec,grec1} has recently emerged as an important paradigm for unifying user modeling and item prediction. Semantic-ID (SID)-based methods~\cite{tiger,wang2024letter,yang2024unifying} encode each item into a short sequence of discrete semantic tokens and formulate recommendation as conditional sequence generation. This provides a compact and structured output space for expressive representation learning and efficient inference~\cite{tiger,wei2026oneloc,deng2025onerec}. Nevertheless, existing SID recommenders~\cite{tiger} are still primarily trained with item-level behavioral supervision and mainly capture behavioral co-occurrence and local transitions~\cite{tiger,wang2024letter,hou2025generating}. Even when textual, collaborative, or multimodal semantics are incorporated into item representations~\cite{liu2024mmgrec,zhai2025multimodal,zhang2026multi}, the training objective does not explicitly separate demand understanding from recommendation decision making. This limitation becomes more pronounced when interactions are sparse, interests evolve rapidly, or future demands cannot be reliably inferred from local patterns~\cite{zhou2019deep,zhou2020s3}.

\begin{figure}[t]
\centering
\includegraphics[width=\linewidth]{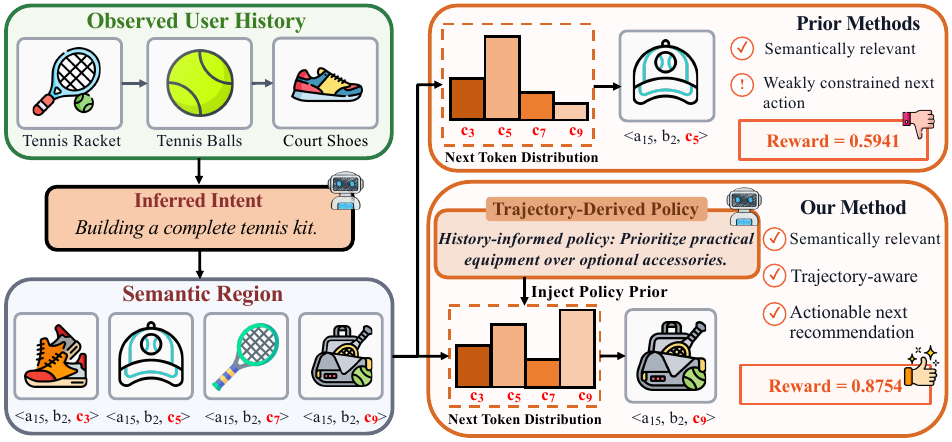}
\caption{
Intent alone provides coarse semantic guidance and is insufficient for fine-grained recommendation, while policy provides actionable decision guidance.
}
\label{fig:motivation}
\end{figure}

Large language models~\cite{chen2024hllm,yi2025recgpt,jia2025learn} offer a promising way to enrich recommendation beyond behavioral co-occurrence. Their world knowledge and semantic reasoning capabilities enable them to connect heterogeneous interactions, recognize functional relations among items, and infer the task behind the user's recent behavior. As shown in Figure~\ref{fig:motivation}, a sequence containing a tennis racket, tennis balls, and court shoes may indicate that the user is assembling a complete tennis kit rather than expressing isolated preferences. Such task-oriented understanding provides a coherent interpretation of the interaction history and identifies the semantic region relevant to the next recommendation, thereby supporting a plausible next-item prediction. However, understanding what the user currently seeks does not necessarily reveal which candidate is most suitable for this particular user.

Within the same intent-consistent region, multiple candidates may be semantically plausible yet differ in their alignment with the user's decision preference. For example, both a tennis cap and a racket bag fit the inferred intent, but a user who favors practical equipment over optional accessories may prefer the racket bag. This motivates two complementary forms of knowledge. \emph{Intent knowledge} identifies what the user currently seeks and constrains the relevant semantic region, while \emph{policy knowledge} guides selection among intent-consistent candidates. A trajectory-derived preference is only a candidate policy by default; it is retained as policy knowledge only when it yields a verified positive gain over an intent-only baseline. Existing LLM-enhanced recommenders~\cite{chen2024hllm,grec2,grec3,grec4,grec5} mainly derive user profiles, semantic preferences, or reasoning signals, but rarely distinguish task understanding from outcome-grounded action selection. We refer to this mismatch as the \emph{Understanding--Action Gap} in generative recommendation.

To bridge this gap, we propose a feedback-driven agent framework for trajectory-grounded policy discovery. A policy agent generates multiple candidates from positive historical trajectories, conditioned on the current intent and behavioral state. A shared executor compares each policy-conditioned recommendation with an intent-only baseline under the same context. Policies are evaluated by their incremental utility, and a feedback agent converts the observed differences into policy-level critiques for iterative refinement. Only policies that yield additional gains are retained.

Although LLMs provide rich intent and policy knowledge, their serving cost hinders online deployment~\cite{cui2024distillation,lin2025can}. We therefore use them only as offline teachers and introduce an Intent Token and a Policy Token into a lightweight SID generator. Their contextualized states form an ordered latent recommendation chain that captures user-specific intent and policy information. Since the language-oriented teacher and behavior-oriented student operate in different representation spaces, we transfer their relational structures rather than directly matching embedding coordinates~\cite{park2019relational, tung2019similarity}. Our dual-space relational distillation aligns first-order user relations and higher-order neighborhood structures in the intent and policy spaces. The resulting model directly infers latent intent and policy states from interaction history and generates the target SID without online LLM inference.

Our main contributions are summarized as follows.

\begin{itemize}
\item We identify the \emph{Understanding--Action Gap} in generative recommendation and introduce outcome-grounded policy knowledge to bridge the gap between intent understanding and recommendation action.

\item We propose a feedback-driven agent framework that discovers and validates policies by their gains over an intent-only baseline, and transfers intent and policy knowledge into a lightweight SID generator for LLM-free online serving.

\item Experiments on public benchmarks show consistent improvements, while online A/B tests achieve gains of 4.506\% in Revenue and 4.621\% in ADVV.
\end{itemize}

\section{Related Work}

\subsection{SID-Based Generative Recommendation}
Generative recommendation formulates item prediction as sequence generation, providing a unified alternative to conventional retrieval and ranking pipelines. Early methods represent each item with a sequence of discrete semantic codes and autoregressively generate the Semantic ID of the target item~\cite{tiger}. Subsequent studies improve Semantic-ID construction and generation from different perspectives. LETTER~\cite{wang2024letter} incorporates semantic information, collaborative signals, and code-assignment regularization into item tokenization, while LIGER~\cite{yang2024unifying} combines generative and dense retrieval to improve robustness. More recent approaches explore end-to-end generative architectures~\cite{deng2025onerec}, longer or parallel SID generation~\cite{hou2025generating}, and recommendation-aware item tokenization. Despite these advances, existing methods remain primarily supervised by the final target item and directly learn the mapping from user history to the target SID. Consequently, demand understanding and recommendation decision making remain entangled in a single item-generation objective, without explicit variables or supervision for either stage. In contrast, our work introduces intent and policy as two complementary variables preceding SID generation.

\subsection{LLM-Enhanced User Understanding}

LLMs have been increasingly used to enhance semantic understanding and reasoning in recommendation. One line of research extracts, summarizes, or dynamically updates user profiles and intents from interaction histories~\cite{bang2025llm,wang2025lettingo,xu2025enhancing}. Agent-based approaches further incorporate tool use, memory, and personalized skills. For example, iAgent~\cite{xu2025iagent} models user instructions through individualized memory, while SAGER~\cite{tao2026sager} extends personalization from user knowledge to user-specific policy skills. Another line introduces explicit or latent reasoning before recommendation. STREAM-Rec~\cite{zhang2025slow} and R2Rec~\cite{zhao2025reason} construct multi-step reasoning traces to support item prediction, whereas Think Before Recommend~\cite{tang2026think}, LARES~\cite{liu2025lares}, and LatentR$^3$~\cite{zhang2025reinforced} perform additional computation in latent space to reduce textual reasoning overhead. OneRec-Think~\cite{liu2025onerec} further integrates explicit reasoning into industrial generative recommendation. These methods mainly enrich user understanding or directly improve item prediction. Even when user-specific skills or policies are introduced, they are typically inferred from user information without recommendation outcome feedback, and thus remain extensions of intent modeling rather than outcome-grounded policy knowledge. In contrast, we separate \emph{intent knowledge} from \emph{policy knowledge}, retaining the latter only when it consistently improves over an intent-only baseline.

\section{Methodology}

\begin{figure*}
    \centering
    \includegraphics[width=1\linewidth]{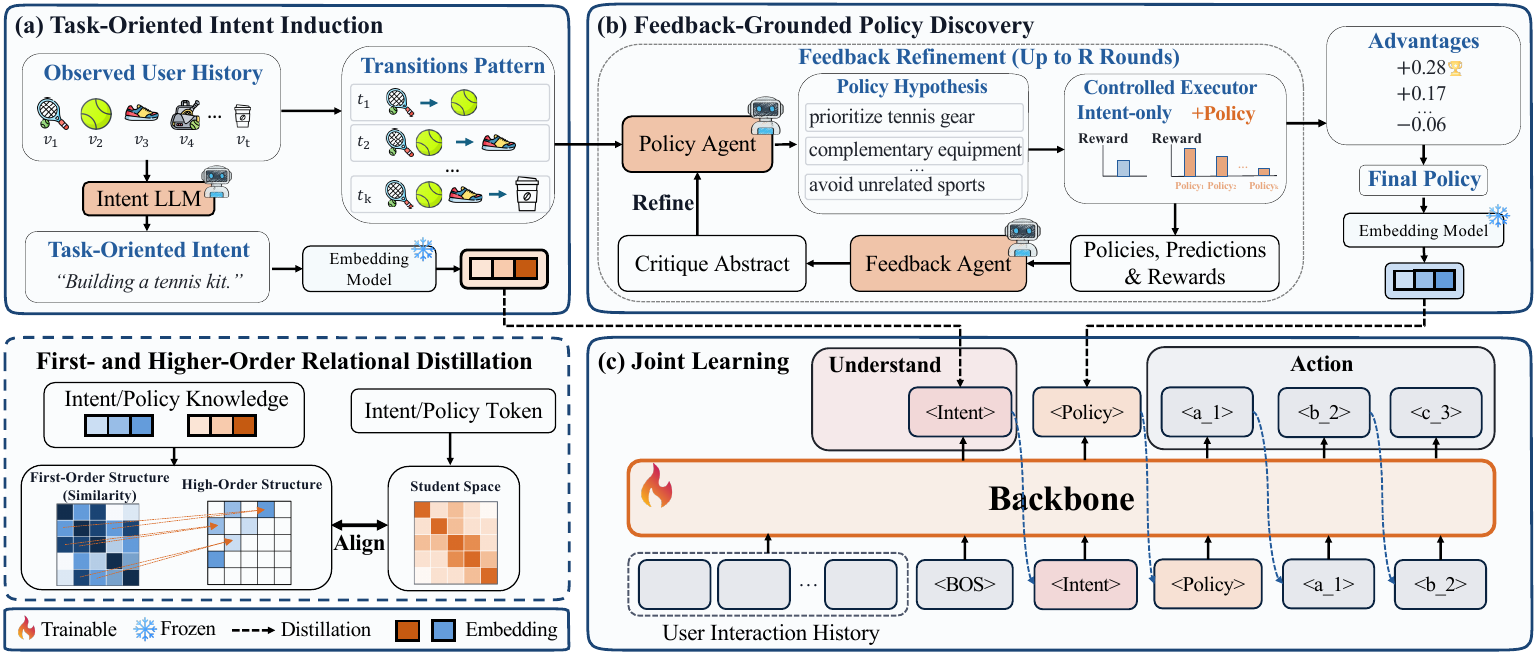}
    \caption{Overview of our framework for bridging the \emph{Understanding--Action Gap}. Task-oriented intent captures what the user currently needs. Policy knowledge determines how the recommender should act. We generate policy hypotheses from historical trajectories and refine them through group-wise feedback. Each policy is evaluated by its gain over an intent-only baseline. Finally, we transfer intent and policy knowledge into ordered Intent and Policy Tokens through dual-space relational distillation. This enables LLM-free online serving.
    }
    \label{fig:placeholder}
\end{figure*}

\subsection{Problem Formulation}

Let $\mathcal{U}$ and $\mathcal{V}$ denote the sets of users and items, respectively. For each user $u\in\mathcal{U}$, we observe a chronological interaction sequence $\mathcal{H}_u=(v_{u,1},\ldots,v_{u,T_u})$, where $T_u$ is the sequence length, and aim to predict the next interacted item $y_u\in\mathcal{V}$. SID-based generative recommendation represents each item $v$ as an $L$-level sequence of discrete semantic codes: 
\begin{equation} \operatorname{SID}(v) = \left( c_v^{(1)}, c_v^{(2)}, \ldots, c_v^{(L)} \right), 
\label{eq:sid_definition} 
\end{equation} 
where $c_v^{(\ell)}$ denotes the code assigned to item $v$ at the $\ell$-th quantization level. Given the interaction history, the target SID is generated autoregressively as 
\begin{equation} p_{\theta} \left( \operatorname{SID}(y_u) \mid \mathcal{H}_u \right) = \prod_{\ell=1}^{L} p_{\theta} \left( c_{y_u}^{(\ell)} \mid \mathcal{H}_u, \mathbf{c}_{y_u}^{(<\ell)} \right), \label{eq:generative_rec} \end{equation} 
where $\mathbf{c}_{y_u}^{(<\ell)} =(c_{y_u}^{(1)},\ldots,c_{y_u}^{(\ell-1)})$ denotes the previously generated SID prefix. Although this formulation provides an efficient framework for next-item generation, its supervision is still derived solely from the target item. The model must therefore implicitly learn both what the user currently seeks and how the recommender should act from the same item-level objective. We refer to this mismatch as the \emph{Understanding--Action Gap}.

To explicitly model these two complementary factors, we introduce a task-oriented intent $I_u$ and a recommendation policy $P_u$. The intent captures the dominant short-term demand underlying the next interaction, while the policy specifies how the recommender should respond to that demand, including a preferred recommendation direction and a rejection boundary. 
Specifically, $I_u$ is inferred from $\mathcal{H}_u$, and $P_u$ is further derived conditioned on both $\mathcal{H}_u$ and $I_u$.


\subsection{Task-Oriented Intent Induction}

We first induce a task-oriented intent to make the user's underlying demand explicit. Unlike a general preference profile that characterizes overall interests, the task-oriented intent captures the demand most relevant to the current recommendation context by jointly considering long-term preferences and recent behavioral evidence. Given the interaction history and the textual metadata of historical items, an LLM-based intent agent $\mathcal{A}_{\mathrm{int}}$ generates a textual intent $I_u$, which is further encoded into a continuous teacher representation: \begin{equation} I_u = \mathcal{A}_{\mathrm{int}} \bigl( \mathcal{H}_u, \mathcal{M}(\mathcal{H}_u) \bigr), \quad \mathbf{e}_u^{I} = \mathcal{E}(I_u). \label{eq:intent_induction} 
\end{equation} 
Here, $\mathcal{M}(\mathcal{H}_u)$ denotes the textual metadata of historical items, and $\mathbf{e}_u^{I}\in\mathbb{R}^{d}$ is the resulting intent teacher representation produced by a frozen semantic encoder $\mathcal{E}$. The intent agent is instructed to infer the demand underlying the observed behavior rather than directly predict a specific item. The resulting intent $I_u$ provides the semantic condition for subsequent policy discovery, while $\mathbf{e}_u^{I}$ serves as the teacher signal for intent knowledge transfer. 

\subsection{Feedback-Grounded Policy Discovery}
\label{sec:policy_discovery}

Intent captures what the user currently seeks, but does not uniquely determine how the recommender should act. Multiple policy hypotheses may appear plausible under the same intent, yet only some provide additional predictive value. To discover outcome-effective policies, we develop a dual-agent framework composed of a \emph{policy agent} and a \emph{feedback agent}, connected through a shared \emph{recommendation executor}. The policy agent proposes and iteratively revises personalized decision principles, while the feedback agent diagnoses their execution failures from recommendation outcomes. The executor provides a controlled environment for evaluating each policy against an intent-only baseline, enabling policy knowledge to be selected and refined according to its incremental recommendation utility rather than linguistic plausibility.

\subsubsection{Policy Agent for Hypothesis Induction}
\label{sec:policy_hypothesis}

The interaction history provides multiple observed prefix–successor transitions. Specifically, for user $u$, we construct $\mathcal{T}_u = \{(\mathcal{H}_u^{<t},v_{u,t})\}_{t=2}^{T_u}$, where $\mathcal{H}_u^{<t} = (v_{u,1},\ldots,v_{u,t-1})$. Each prefix--successor pair provides observational evidence about how the user's demand may translate into a subsequent interaction. Conditioned on these transitions, the historical item metadata, and the inferred intent, the policy agent $\mathcal{A}_{\mathrm{pol}}$ generates $K$ personalized policy hypotheses: \begin{equation}
    \mathcal{P}_u^{(0)}
    =
    \left\{
        P_{u,1}^{(0)},
        \ldots,
        P_{u,K}^{(0)}
    \right\}
    =
    \mathcal{A}_{\mathrm{pol}}
    \left(
        \mathcal{T}_u,
        \mathcal{M}(\mathcal{H}_u),
        I_u
    \right).
    \label{eq:policy_hypotheses}
\end{equation}
Each hypothesis specifies a \emph{recommendation direction} to prioritize and a \emph{rejection boundary} to suppress plausible but contextually unsuitable alternatives. The resulting candidates represent different ways of acting under the same inferred intent, rather than alternative descriptions of the intent itself.

Because observed transitions provide only hypothesis-induction evidence, they do not guarantee that every generated policy improves recommendation. We therefore evaluate all candidates through controlled execution before using them as policy supervision.

\subsubsection{Advantage-Based Policy Evaluation}
\label{sec:policy_validation}

Let $\mathcal{F}_{\mathrm{exec}}$ denote a fixed recommendation executor that generates a textual profile of the predicted next item. Given the same history and intent, the executor first produces an intent-only prediction
$\widehat{M}_{u,0}=\mathcal{F}_{\mathrm{exec}}(\mathcal{H}_u,I_u)$
and a policy-conditioned prediction
$\widehat{M}_{u,k}^{(r)}=\mathcal{F}_{\mathrm{exec}}(\mathcal{H}_u,I_u,P_{u,k}^{(r)})$
for each candidate at evolution round $r$.

Let $M_u^{+}$ denote the textual profile of the last item in the user's training sequence. Using a frozen semantic encoder $\mathcal{E}$, we define
\begin{equation}
\begin{aligned}
R_{u,0}
&=
\operatorname{cos}\left(
\mathcal{E}(\widehat{M}_{u,0}),
\mathcal{E}(M_u^{+})
\right),\\
A_{u,k}^{(r)}
&=
\operatorname{cos}\left(
\mathcal{E}(\widehat{M}_{u,k}^{(r)}),
\mathcal{E}(M_u^{+})
\right)
-
R_{u,0}.
\end{aligned}
\label{eq:policy_advantage}
\end{equation}

The advantage $A_{u,k}^{(r)}$ measures the incremental utility introduced by the policy beyond the information already captured by the history and intent. We use semantic similarity as a dense policy-level reward because rank-based metrics such as Recall@$K$ and NDCG@$K$ are too sparse for iterative policy refinement. When the last item in the user's training sequence remains outside the top-$K$
list, all candidate policies receive zero reward, even if some substantially
improve its retrieval score. Such indistinguishable feedback provides no direction for comparing candidate policies or refining them in subsequent rounds.

Within the controlled policy-evaluation pipeline, predicted item profiles are
matched against item representations using the frozen semantic encoder and
similarity function. The similarity to the last item in the user's training
sequence therefore provides graded feedback before it enters the top-$K$ list. Subtracting the intent-only score further isolates the incremental progress introduced specifically by the policy. Rather than selecting a single candidate immediately, we retain the complete set of hypotheses as comparative evidence for subsequent reflection.

\subsubsection{Feedback Agent for Policy Evolution}
\label{sec:policy_refinement}

At each round, the feedback agent jointly examines the candidate policies,
their execution outcomes, and the corresponding advantages:
\begin{equation}
Z_u^{(r)}
=
\mathcal{A}_{\mathrm{fb}}
\left(
\left\{
P_{u,k}^{(r)},
\widehat{M}_{u,k}^{(r)},
A_{u,k}^{(r)}
\right\}_{k=1}^{K}
\right).
\label{eq:group_policy_feedback}
\end{equation}
By comparing candidates evaluated under the same history, intent, executor,
and reward function, the feedback agent identifies decision patterns
associated with higher rewards, shared failure modes among weaker candidates,
and policy directions that remain insufficiently explored. This group-wise
reflection exploits cross-candidate evidence that cannot be obtained by
independently refining each policy.

We identify the best policy in the current round as
\begin{equation}
k_u^{(r)}
=
\arg\max_{1\leq k\leq K} A_{u,k}^{(r)},
\qquad
P_{u,\mathrm{best}}^{(r)}
=
P_{u,k_u^{(r)}}^{(r)}.
\label{eq:round_best_policy}
\end{equation}
The corresponding best advantage provides a reference for assessing whether
subsequent evolution yields further improvement.

Conditioned on the interaction history, intent, current policy set, and
group-level critique, the policy agent generates a new set of $K$ diverse
hypotheses:
\begin{equation}
\mathcal{P}_u^{(r+1)}
=
\mathcal{A}_{\mathrm{pol}}
\left(
\mathcal{H}_u,
I_u,
\mathcal{P}_u^{(r)},
Z_u^{(r)}
\right).
\label{eq:policy_regeneration}
\end{equation}
The regeneration prompt encourages the agent to preserve effective decision
patterns, correct recurring failures, and explore complementary recommendation
directions and rejection boundaries.

The newly generated candidates are evaluated by the same controlled executor
and evaluation function. Evolution continues only if the best advantage in the
new round exceeds that of the previous round; otherwise, the process terminates.
It also stops after at most $R$ rounds. Finally, we select the highest-advantage
policy encountered throughout the evolution process:
\begin{equation}
(r_u^{*},k_u^{*})
=
\arg\max_{\substack{0\leq r\leq R\\1\leq k\leq K}}
A_{u,k}^{(r)},
\qquad
P_u^{*}
=
P_{u,k_u^{*}}^{(r_u^{*})},
\qquad
\mathbf{e}_u^{P}
=
\mathcal{E}(P_u^{*}).
\label{eq:final_policy_selection}
\end{equation}
Thus, the policy teacher representation
$\mathbf{e}_u^{P}\in\mathbb{R}^{d}$ is derived from the
highest-advantage policy identified during multi-hypothesis evolution,
rather than directly from the initial LLM generation.

Policy discovery and evolution are performed exclusively on training
interactions, and policy supervision is applied only to users with at least one
candidate whose advantage is positive. Users without such candidates remain
available for recommendation learning and intent distillation but are excluded
from teacher-side policy distillation. Validation and test targets are never used
for policy generation, evaluation, or refinement. Overall, the resulting policy
knowledge is grounded in three levels of evidence: historical transitions support
hypothesis generation, positive advantage verifies incremental utility, and
iterative group-wise feedback progressively refines the resulting decision
principles.

\subsection{Latent Intent--Policy Knowledge Transfer}

The intent and policy teachers provide explicit demand and decision knowledge, but invoking them during online inference would incur substantial serving overhead. Moreover, their representations are derived from natural-language descriptions, whereas the student learns behavior-oriented representations for SID generation. Direct feature matching is therefore unnecessarily restrictive, as the teacher and student spaces need not share the same coordinates. To address these challenges, we use the LLM agents only for offline supervision and transfer their relational knowledge into two latent states of a lightweight SID generator. Specifically, we introduce a \emph{latent intent--policy chain} into the student and align its user-level \emph{relational structures} with the corresponding teacher spaces.



\subsubsection{Latent Intent--Policy Recommendation Chain}
\label{sec:progressive_chain}


To internalize intent and policy knowledge within the generative recommender, we extend the original SID decoding sequence with two latent tokens: an Intent Token $q^{I}$ and a Policy Token $q^{P}$. 
Given the encoded interaction history $\mathcal{H}_u$, the decoder first processes the Intent Token, then the Policy Token, and finally generates the target SID codes autoregressively. We denote the decoder hidden states at the Intent and Policy Token positions as $\mathbf{h}_u^{I}$ and $\mathbf{h}_u^{P}$, respectively. Since the Intent Token is inferred directly from the behavioral history, $\mathbf{h}_u^{I}$ captures the user's underlying demand. The Policy Token is subsequently conditioned on both the history and the preceding Intent Token, allowing $\mathbf{h}_u^{P}$ to represent how the recommender should act under that
demand. The target SID is then generated conditioned on both latent tokens, organizing recommendation into a latent \emph{understand--plan--generate} chain.

\subsubsection{First- and Higher-Order Relational Distillation}
\label{sec:relational_distillation}

The LLM-derived intent and policy representations reside in semantic spaces that are not directly aligned with the behavior-oriented latent space of the recommender. We therefore transfer their user-level relational structures rather than matching individual embeddings. For knowledge type $X\in\{I,P\}$, let $\mathbf{e}_u^X$ denote the teacher representation and $\mathbf{z}_u^X=g_X(\mathbf{h}_u^X)$ the projected student representation. We use the same relational objective for intent and policy, while policy distillation is applied only to users with validated policy supervision.

\paragraph{First-order relations.}
We first preserve the direct relations between users. For
$\nu\in\{\mathrm{T},\mathrm{S}\}$, let
$\mathbf{r}_u^{X,\mathrm{T}}=\mathbf{e}_u^X$ and
$\mathbf{r}_u^{X,\mathrm{S}}=\mathbf{z}_u^X$.
The relational distribution of user $i$ in the batch is
defined as
\begin{equation}
\pi_{ij}^{X,\nu}
=
\frac{
\exp\!\left(
\operatorname{sim}(\mathbf{r}_i^{X,\nu},\mathbf{r}_j^{X,\nu})/\tau_\nu
\right)
}{
\sum_{m\in\mathcal{B}_X\setminus\{i\}}
\exp\!\left(
\operatorname{sim}(\mathbf{r}_i^{X,\nu},\mathbf{r}_m^{X,\nu})/\tau_\nu
\right)
},
\quad
j\in\mathcal{B}_X\setminus\{i\}.
\label{eq:first_order_relation}
\end{equation}
The teacher and student distributions are defined over the same user set,
making their direct relational structures comparable.

\paragraph{Higher-order relations.}
First-order relations describe direct user proximity but do not capture whether
users exhibit similar relational patterns. We therefore retain the teacher-defined
top-$K_1$ entries of each first-order distribution to construct a sparse
neighborhood profile $\widetilde{\boldsymbol{\pi}}_i^{X,\nu}$, using the same
indices for the teacher and student. The higher-order affinity is defined as
\begin{equation}
g_{ij}^{X,\nu}
=
\left\langle
\widetilde{\boldsymbol{\pi}}_i^{X,\nu},
\widetilde{\boldsymbol{\pi}}_j^{X,\nu}
\right\rangle.
\label{eq:higher_order_affinity}
\end{equation}
For each user, we exclude itself and its first-order top-$K_1$ neighbors,
retain the top-$K_2$ remaining users according to $g_{ij}^{X,\mathrm{T}}$,
and apply a row-wise softmax to obtain the higher-order distribution
$\boldsymbol{\rho}_i^{X,\nu}$. This captures structurally similar users beyond
the immediate neighborhood without repeatedly transferring the same direct
relations.

The two relational levels are jointly transferred through 
\begin{equation}
\begin{aligned}
\mathcal{L}_{X}^{\mathrm{rel}}
=
\frac{1}{|\mathcal{B}_X|}
\sum_{i\in\mathcal{B}_X}
\Bigg[
&\operatorname{KL}
\left(
\operatorname{sg}
[\boldsymbol{\pi}_i^{X,\mathrm{T}}]
\,\middle\|\,
\boldsymbol{\pi}_i^{X,\mathrm{S}}
\right) \\
&+
\gamma\,
\operatorname{KL}
\left(
\operatorname{sg}
[\boldsymbol{\rho}_i^{X,\mathrm{T}}]
\,\middle\|\,
\boldsymbol{\rho}_i^{X,\mathrm{S}}
\right)
\Bigg],
\quad
X\in\{I,P\}.
\end{aligned}
\label{eq:relational_distillation}
\end{equation}
where $\operatorname{sg}[\cdot]$ denotes stop-gradient and $\gamma$ controls the contribution of higher-order transfer. The first-order term preserves direct semantic neighborhoods, while the higher-order term transfers complementary structural relations beyond those neighbors.

\subsubsection{Joint Learning and Inference}
\label{sec:behavior_anchored_learning}

Relational distillation structures the Intent and Policy Token spaces but does not directly supervise next-item prediction. We therefore combine it with the autoregressive recommendation objective:
\begin{equation} 
\mathcal{L}_{\mathrm{rec}} = - \frac{1}{|\mathcal{B}|} \sum_{u\in\mathcal{B}} \sum_{\ell=1}^{L} \log p_{\theta} \left( c_{y_u}^{(\ell)} \,\middle|\, \mathcal{H}_u, q^{I}, q^{P}, \mathbf{c}_{y_u}^{(<\ell)} \right), \label{eq:recommendation_loss} 
\end{equation} 
where $\mathbf{c}_{y_u}^{(<\ell)} = (c_{y_u}^{(1)},\ldots,c_{y_u}^{(\ell-1)})$ denotes the previously generated SID prefix. 
The complete objective is 
\begin{equation} \mathcal{L} = \mathcal{L}_{\mathrm{rec}} + \lambda_I \mathcal{L}_{I}^{\mathrm{rel}} + \lambda_P \mathcal{L}_{P}^{\mathrm{rel}}, \label{eq:joint_objective} 
\end{equation} 
where $\lambda_I$ and $\lambda_P$ control the strengths of intent and policy transfer, respectively. The recommendation objective directly optimizes the latent tokens for next-item prediction, while the intent and policy distillation objectives introduce complementary demand-level and decision-level supervision. 

During inference, only the SID recommender and the two latent token positions are retained. The LLM agents, textual intent and policy descriptions, semantic encoder, and projection heads are used exclusively for offline supervision and introduce no online serving cost. The student directly infers the Intent and Policy Token states from the interaction history before generating the target SID.

\section{Experiments}

In this section, we conduct extensive experiments to evaluate the proposed framework through the following research questions: \\
\textbf{RQ1: Effectiveness.} How does our method compare with strong baselines in offline and online settings?\\
\textbf{RQ2: Understanding--Action Gap.} Does policy knowledge complement intent understanding?\\
\textbf{RQ3: Policy Discovery.} Does feedback-driven policy discovery outperform direct generation and random selection?\\
\textbf{RQ4: Knowledge Distillation.}
How effectively does first- and higher-order relational distillation transfer intent and policy knowledge to the lightweight student?\\
\textbf{RQ5: Efficiency.} What effectiveness--efficiency trade-off does the proposed method achieve in online serving?



\subsection{Experimental Setting}

\subsubsection{Datasets}

We conduct experiments on three widely used categories from the Amazon Review
dataset~\cite{mcauley2015image}: \emph{Beauty}, \emph{Toys and Games}, and
\emph{Sports and Outdoors}. Detailed preprocessing procedures and dataset statistics are provided in Appendix~\ref{app:dataset_details}.



\begin{table*}[t]
    \centering
    \caption{
    Overall performance comparison on three Amazon datasets.
    }
    \label{tab:overall_performance}
    \begin{tabular}{lcccccccccccc}
        \toprule
        \multirow{2}{*}{\textbf{Method}}
        & \multicolumn{4}{c}{\textbf{Beauty}}
        & \multicolumn{4}{c}{\textbf{Toys and Games}}
        & \multicolumn{4}{c}{\textbf{Sports and Outdoors}} \\
        \cmidrule(lr){2-5}
        \cmidrule(lr){6-9}
        \cmidrule(lr){10-13}
        & R@5 & R@10 & N@5 & N@10
        & R@5 & R@10 & N@5 & N@10
        & R@5 & R@10 & N@5 & N@10 \\
        \midrule
        P5
        & 0.0163 & 0.0254 & 0.0107 & 0.0136
        & 0.0070 & 0.0121 & 0.0050 & 0.0066
        & 0.0061 & 0.0095 & 0.0041 & 0.0052 \\

        Caser
        & 0.0205 & 0.0347 & 0.0131 & 0.0176
        & 0.0166 & 0.0270 & 0.0107 & 0.0141
        & 0.0116 & 0.0194 & 0.0072 & 0.0097 \\

        GRU4Rec
        & 0.0164 & 0.0283 & 0.0099 & 0.0137
        & 0.0097 & 0.0176 & 0.0059 & 0.0084
        & 0.0129 & 0.0204 & 0.0086 & 0.0110 \\

        SASRec
        & 0.0387 & 0.0605 & 0.0249 & 0.0318
        & 0.0463 & 0.0675 & 0.0306 & 0.0374
        & 0.0233 & 0.0350 & 0.0154 & 0.0192 \\

        BERT4Rec
        & 0.0203 & 0.0347 & 0.0124 & 0.0170
        & 0.0116 & 0.0203 & 0.0071 & 0.0099
        & 0.0115 & 0.0191 & 0.0075 & 0.0099 \\

        HGN
        & 0.0325 & 0.0512 & 0.0206 & 0.0266
        & 0.0321 & 0.0497 & 0.0221 & 0.0277
        & 0.0189 & 0.0313 & 0.0120 & 0.0159 \\

        \midrule

        LLM
        & 0.0293 & 0.0413 & 0.0203 & 0.0241
        & 0.0416 & 0.0645 & 0.0278 & 0.0352
        & 0.0146 & 0.0233 & 0.0089 & 0.0118 \\

        \midrule

        TIGER
        & 0.0425 & 0.0617 & 0.0279 & 0.0333
        & 0.0336 & 0.0510 & 0.0217 & 0.0273
        & 0.0216 & 0.0332 & 0.0140 & 0.0177 \\

        \textbf{TIGER + Ours}
        & \textbf{0.0491} & \textbf{0.0739}
        & \textbf{0.0341} & \textbf{0.0417}
        & \textbf{0.0485} & \textbf{0.0712}
        & \textbf{0.0322} & \textbf{0.0393}
        & \textbf{0.0288} & \textbf{0.0441}
        & \textbf{0.0186} & \textbf{0.0234} \\

        \midrule

        LETTER
        & 0.0447 & 0.0693 & 0.0296 & 0.0375
        & 0.0366 & 0.0542 & 0.0239 & 0.0296
        & 0.0232 & 0.0372 & 0.0143 & 0.0188 \\

        \textbf{LETTER + Ours}
        & \textbf{0.0513} & \textbf{0.0753}
        & \textbf{0.0347} & \textbf{0.0423}
        & \textbf{0.0497} & \textbf{0.0724}
        & \textbf{0.0345} & \textbf{0.0415}
        & \textbf{0.0282} & \textbf{0.0446}
        & \textbf{0.0183} & \textbf{0.0233} \\

        \bottomrule
    \end{tabular}
\end{table*}




\subsubsection{Baselines}

We compare our method against six conventional recommenders
(Caser~\cite{tang2018caser}, GRU4Rec~\cite{hidasi2016gru4rec}, SASRec~\cite{kang2018self}, BERT4Rec~\cite{sun2019bert4rec}, HGN~\cite{ma2019hgn}, and P5~\cite{geng2022recommendation}), two SID-based generative
recommenders (TIGER~\cite{tiger} and LETTER~\cite{wang2024letter}), and a direct LLM baseline based on
Qwen3.5-122B-A10B~\cite{yang2025qwen3}. Detailed descriptions of all baselines are provided in
Appendix~\ref{app:baseline_details}.



\subsubsection{Evaluation Metrics}

We evaluate recommendation performance using $\mathrm{Recall}@K$~\cite{sun2019bert4rec} and $\mathrm{NDCG}@K$~\cite{jarvelin2002cumulated}, where $K\in\{5,10\}$. Under the leave-one-out setting, $\mathrm{Recall}@K$ measures whether the held-out target item appears in the top-$K$ recommendation list, while $\mathrm{NDCG}@K$ further considers its ranking position and assigns higher scores to items ranked closer to the top. Higher values indicate better recommendation performance.
For the online A/B test, we use \emph{Advertiser Value} (ADVV) and \emph{Revenue} as the primary business metrics. ADVV measures the amount that the platform can reasonably charge advertisers for the delivered traffic, while Revenue denotes the platform's realized income. The desired outcome is to improve both metrics while maintaining a larger relative increase in ADVV than in Revenue.



\subsubsection{Implementation Details}

For all SID-based methods, item text is constructed by concatenating the title and
category and encoded with Qwen3-Embedding-8B~\cite{zhang2025qwen3embedding}.
We use TIGER~\cite{tiger} and LETTER~\cite{wang2024letter} as two generative backbones, and instantiate the intent and policy agents with
Qwen3.5-122B-A10B~\cite{yang2025qwen3}. 
Additional training and baseline details are
provided in Appendix~\ref{app:implementation_details}.

\subsection{Main Results}

Table~\ref{tab:overall_performance} reports the overall results on three Amazon datasets. The proposed method consistently outperforms both conventional sequential recommendation models and existing SID-based generative methods, demonstrating the effectiveness of explicitly modeling intent and policy knowledge for next-item generation. Moreover, applying our framework to different backbones leads to consistent improvements across datasets and evaluation metrics. These results verify that our method is not tied to a specific architecture and can be readily integrated into different SID-based generative recommendation backbones.


\subsection{Model Analysis}

\subsubsection{Analysis of the Understanding--Action Gap (RQ2)}
\label{sec}

We examine whether intent understanding alone is sufficient for recommendation and whether policy provides complementary decision knowledge. All variants retain the Intent Token and intent distillation, differing only in
the use of the Policy Token and policy distillation. Experiments are conducted on Beauty with TIGER \cite{tiger} as the backbone; results on the other datasets are reported in Appendix~\ref{app:additional_dataset_results}. 
On Beauty, $87.74\%$ of the training users have at least one candidate policy
with a positive advantage over the intent-only baseline and therefore receive
policy supervision.
As shown in Table~\ref{tab:rq2_component_ablation}, adding the Policy Token
outperforms the intent-only setting, indicating that intent does not fully
determine the final item-level decision. The gain remains limited without
policy distillation, suggesting that the improvement is not merely due to an
additional latent position. 
In contrast, the full model achieves the best performance by transferring validated policy knowledge, providing empirical evidence for the proposed \emph{Understanding--Action Gap}.

\begin{table}[t]
    \centering
    \small
    \setlength{\tabcolsep}{3pt}
    \caption{
    Effect of policy token and policy distillation on Beauty with TIGER. All variants retain the Intent Token and intent distillation.
    }
    \label{tab:rq2_component_ablation}
    \begin{tabular}{cc|cccc}
        \toprule
        \textbf{Policy Token}
        & \textbf{Policy Distillation}
        & \textbf{R@5}
        & \textbf{R@10}
        & \textbf{N@5}
        & \textbf{N@10} \\
        \midrule
        -- & --
        & 0.0448
        & 0.0682
        & 0.0307
        & 0.0383 \\
        \checkmark & --
        & 0.0478 & 0.0722 & 0.0325 & 0.0401 \\
        \checkmark & \checkmark
        & \textbf{0.0491}
        & \textbf{0.0739}
        & \textbf{0.0341}
        & \textbf{0.0417} \\
        \bottomrule
    \end{tabular}
\end{table}



Table~\ref{tab:rq2_sid_hierarchy} further reveals a coarse-to-fine division of
labor. We use $\mathrm{R}_{1:\ell}@10$ to denote Recall@10 measured by whether
the predicted SID matches the target prefix up to level $\ell$. The reported
values are absolute improvements in percentage points. Intent brings its
largest gain at the first SID level, indicating better localization of the
target's broad semantic region. Policy contributes little at the first level
but yields larger gains at deeper levels, where finer-grained item
discrimination is required. This pattern is consistent with their respective
roles: intent identifies the demand region, whereas policy resolves the action
within that region.

\begin{table}[t]
\centering
\small
\caption{
Stage-wise contributions of intent and policy on Beauty with TIGER as the backbone.
$\mathrm{R}_{1:\ell}@10$ denotes SID-prefix Recall@10 up to level $\ell$;
values are absolute gains in percentage points.
}
\label{tab:rq2_sid_hierarchy}
\begin{tabular}{lccc}
\toprule
\textbf{Model Comparison}
& $\boldsymbol{\Delta \mathrm{R}_{1:1}@10}$
& $\boldsymbol{\Delta \mathrm{R}_{1:2}@10}$
& $\boldsymbol{\Delta \mathrm{R}_{1:3}@10}$ \\
\midrule
TIGER $\rightarrow$ TIGER + Intent
& \textbf{+2.23}
& +1.63
& +1.19 \\
TIGER + Intent $\rightarrow$ Full Model
& +0.01
& +0.18
& \textbf{+0.37} \\
\bottomrule
\end{tabular}
\end{table}

\subsubsection{Analysis of Feedback-Grounded Policy Discovery (RQ3)}
\label{sec:policy_discovery_analysis}

We investigate whether effective policies require advantage-grounded selection
and iterative refinement, and whether their benefits go beyond arbitrary
LLM-generated policies or direct target-semantic supervision. Experiments are
conducted on Beauty with TIGER~\cite{tiger} as the backbone, while results on
the other datasets are reported in
Appendix~\ref{app:additional_dataset_results}. We compare the full model with
three variants. \emph{w/o Policy Evolution} selects the highest-advantage
policy from the initial candidate set without further refinement.
\emph{Random Policy Selection} randomly chooses an initial policy without
advantage-based validation. \emph{Target-as-Policy} replaces the discovered
policy with the target-item description while retaining the same Policy Token
and distillation pipeline. As shown in
Table~\ref{tab:rq3_evolution_ablation}, removing policy evolution consistently
degrades performance, while random selection performs even worse. These
results confirm the importance of both advantage-grounded validation and
iterative refinement. The inferior performance of Target-as-Policy further
shows that the gains arise from abstract decision principles rather than direct
injection of target semantics.

\begin{figure}[t]
    \centering
    \includegraphics[scale=0.4]{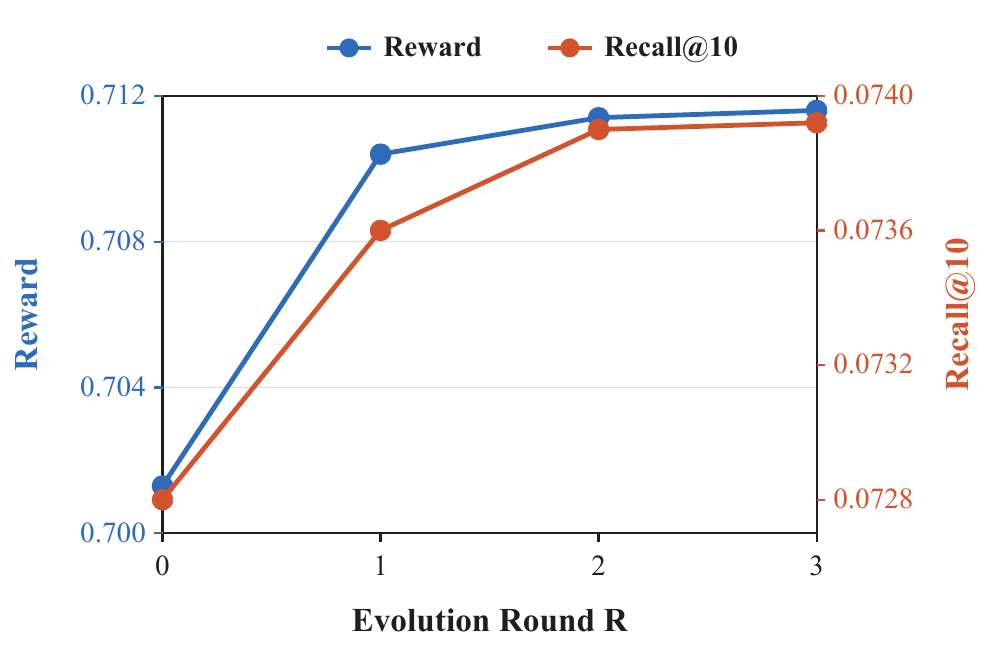}
    \caption{Effect of policy-evolution rounds on the average best-policy
    similarity reward and Recall@10.}
    \label{fig:evolution_round_analysis}
\end{figure}

We examine policy evolution across feedback rounds. At each round, the
best policy provides a reference for assessing whether the next set of $K$
hypotheses yields improvement. As shown in
Figure~\ref{fig:evolution_round_analysis}, both the similarity reward and
Recall@10 increase consistently, with most gains achieved within the first two
rounds. Unlike sparse rank-based metrics, which assign zero feedback whenever the last
item in the user's training sequence remains outside the top-$K$ list, similarity
provides graded signals under the same semantic evaluation function used for
policy comparison. Their consistent trends show that this reward can guide
policy refinement while remaining aligned with recommendation performance.
\begin{table}[t]
    \centering
    \caption{
    Ablation of policy selection and evolution on Beauty with TIGER as the backbone.
    }
    \label{tab:rq3_evolution_ablation}
    \begin{tabular}{lcccc}
        \toprule
        \textbf{Policy Construction}
        & \textbf{R@5}
        & \textbf{R@10}
        & \textbf{N@5}
        & \textbf{N@10} \\
        \midrule
        \textbf{Full Model}
        & \textbf{0.0491}
        & \textbf{0.0739}
        & \textbf{0.0341}
        & \textbf{0.0417} \\
        w/o Policy Evolution
        & 0.0483
        & 0.0728
        & 0.0331
        & 0.0407 \\
        Random Policy Selection
        & 0.0470
        & 0.0718
        & 0.0319
        & 0.0398 \\
        Target-as-Policy
        & 0.0474
        & 0.0703
        & 0.0314
        & 0.0385 \\
        \bottomrule
    \end{tabular}
\end{table}

\subsubsection{Analysis of Knowledge Distillation (RQ4)}
\label{sec:rq4}

We evaluate how effectively intent and policy knowledge is transferred to the lightweight student model. As shown in Table~\ref{tab:distillation}, removing either intent or policy distillation degrades performance, while removing both leads to the largest overall drop. This confirms that the two supervision signals provide complementary demand-level and decision-level knowledge beyond the standard next-item objective. 
The larger degradation without intent distillation
suggests that intent provides the primary semantic foundation, which policy
further refines. Removing higher-order distillation also hurts performance,
showing that neighborhood-level structure complements direct user relations.
Moreover, direct MSE matching is consistently inferior to relational
distillation, supporting the use of relation-level alignment across
heterogeneous teacher and student spaces.

\begin{figure}[t]
    \centering
    \includegraphics[width=\linewidth]{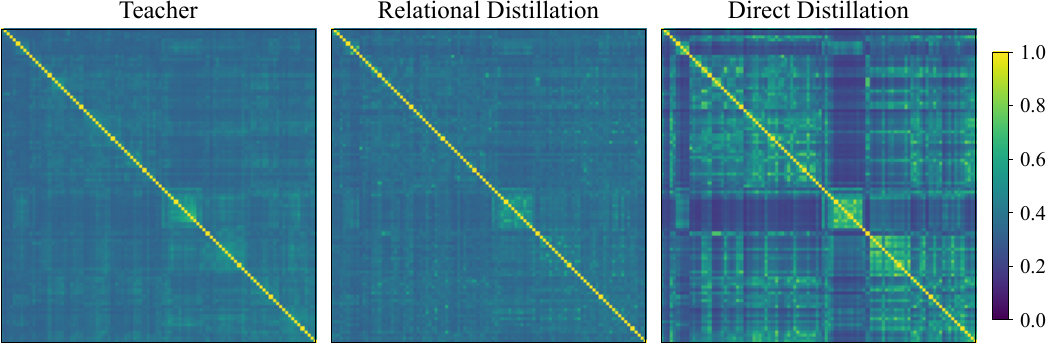}
    \caption{
   Pairwise similarity structures. From left to right: the teacher policy space, our relationally distilled Policy Token space, and the directly distilled Policy Token space.
    }
    \label{fig:policy_structure}
\end{figure}

Figure~\ref{fig:policy_structure} further compares the pairwise similarity structures of the teacher policy representations and the learned Policy Tokens. Relational distillation preserves a structure closer to the teacher, whereas direct MSE alignment introduces more pronounced block-wise distortions. This structural fidelity is also reflected in its stronger recommendation
performance. The corresponding intent-space visualization is provided in the Appendix~\ref{app:intent_structure_visualization}.

\begin{table}[t]
    \centering
    \caption{Ablation study of knowledge distillation on Beauty.}
    \label{tab:distillation}
    \setlength{\tabcolsep}{2.5pt}
    \begin{tabular}{lcccc}
        \toprule
        Method & R@5 & R@10 & N@5 & N@10 \\
        \midrule
        \textbf{Relational Distillation (Ours)} & \textbf{0.0491} & \textbf{0.0739}
                      & \textbf{0.0341} & \textbf{0.0417} \\
        Direct Distillation & 0.0469 & 0.0698 & 0.0318 & 0.0391 \\
        w/o Higher-Order Distillation & 0.0481 & 0.0721 & 0.0330 & 0.0406 \\
        w/o Intent Distillation & 0.0466 &  0.0711 & 0.0311 &  0.0389  \\
        w/o Policy Distillation & 0.0478 & 0.0722 & 0.0325 & 0.0401 \\
        w/o Distillation  & 0.0457 & 0.0709 & 0.0308 & 0.0388 \\

        \bottomrule
    \end{tabular}
\end{table}

\subsubsection{Analysis of Hyperparameter Sensitivity}

\begin{figure}[t]
    \centering
    \includegraphics[width=0.49\linewidth]
    {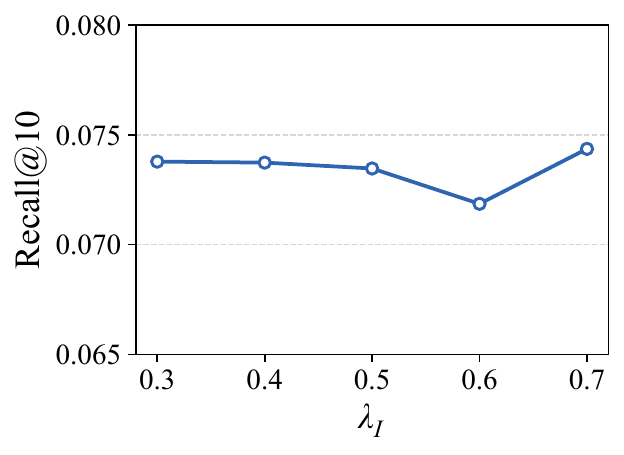}
    \hfill
    \includegraphics[width=0.49\linewidth]
    {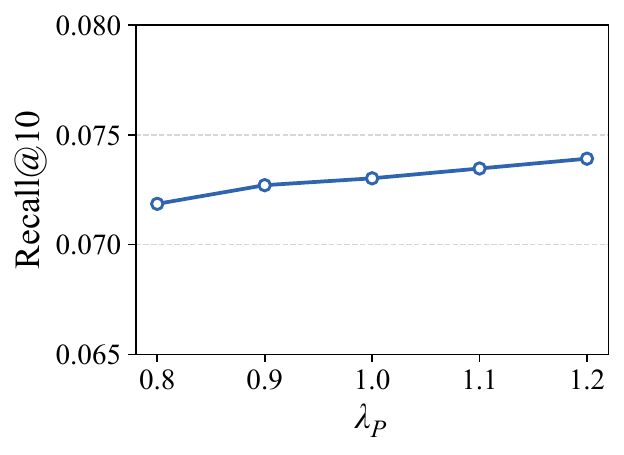}
    \caption{
        Sensitivity of Recall@10 to the intent and policy distillation weights
        on Beauty with TIGER as the backbone.
    }
    \label{fig:distill_weight_sensitivity}
\end{figure}

We study the sensitivity of the intent and policy distillation weights, $\lambda_I$ and $\lambda_P$, on Beauty with TIGER as the backbone. As shown in Figure~\ref{fig:distill_weight_sensitivity}, the performance remains relatively stable, demonstrating that our method is not sensitive to these hyperparameters.

\subsection{Online Deployment}
\label{sec:online_deployment}

\subsubsection{Deployment Pipeline}

\begin{figure}[t]
    \centering
    \includegraphics[width=\linewidth]{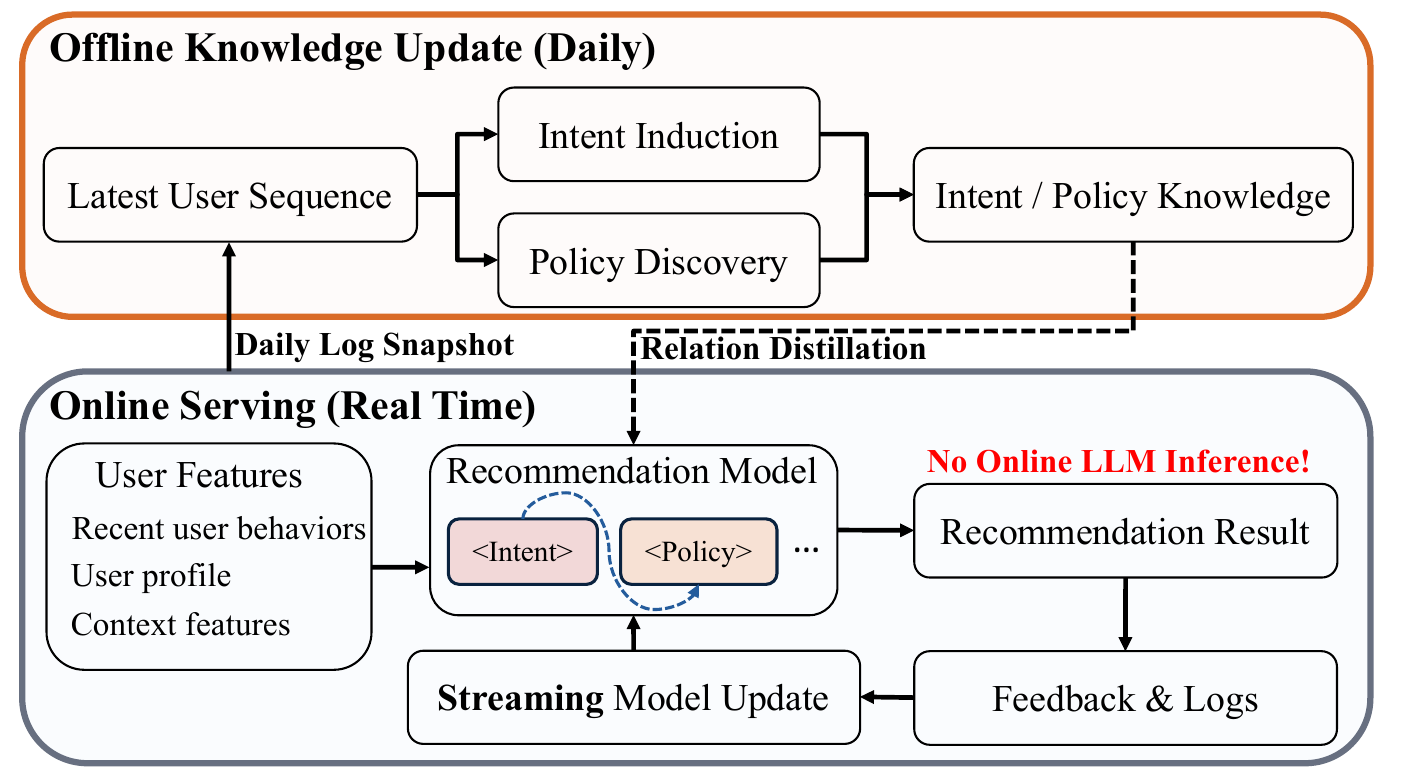}
    \caption{
 Online Deployment.
    }
    \label{fig:online_deployment}
\end{figure}

We deploy the proposed framework in Kuaishou's local-services advertising system using a multi-timescale architecture, as illustrated in Figure~\ref{fig:online_deployment}. The LLM teacher generates user-level intent and policy supervision once per day, while the lightweight SID generator is continuously refreshed through streaming training and performs real-time SID generation. Using completed historical interactions, the LLM teacher induces task-oriented intents and discovers validated recommendation policies. These intent and policy signals remain fixed during the day and serve as teacher supervision for relational distillation. Meanwhile, the SID generator is updated with newly observed interactions. For each online request, the student model takes real-time behavioral and contextual features as input, forms latent Intent Token and Policy Token states, and generates the target SID for candidate retrieval and ranking. This design combines daily LLM knowledge refresh, continuous student-model updates, and real-time SID generation, without invoking the LLM in the online serving path.

\subsubsection{Online A/B Test}

We conduct a seven-day online A/B test in Kuaishou's local-services advertising system, covering approximately $13.25$ million users and $1.61$ million items. The control and treatment groups each receive 5\% of the total traffic, yielding about $70$ million interaction samples during the experiment. Compared with the production baseline, the proposed method improves Revenue by $4.506\%$ and ADVV by $4.621\%$. The larger gain in ADVV indicates that the model not only increases platform revenue but also delivers greater value to advertisers. Both improvements are statistically significant at the 95\% confidence level. These results demonstrate the practical effectiveness of the proposed framework in a large-scale industrial environment.

\subsubsection{Serving Efficiency (RQ5)}
\label{sec:rq5}

A potential concern is that the latent intent--policy chain introduces two additional generation steps during serving. We therefore compare the average per-sample inference latency under the same serving configuration. As shown in Table~\ref{tab:inference_latency}, introducing the two latent tokens increases latency only slightly, from $0.023$ to $0.032$ seconds per sample. In contrast, direct LLM inference is over two orders of magnitude slower. This shows that the proposed framework preserves the efficiency of the base online model while effectively internalizing LLM-derived intent and policy knowledge.

\begin{table}[t]
    \centering
    \setlength{\belowcaptionskip}{-5pt}
    \caption{
    Average per-sample inference latency under the same serving configuration.
    }
    \label{tab:inference_latency}
    \begin{tabular}{lc}
        \toprule
        \textbf{Method}
        & \textbf{Latency (s/sample)} \\
        \midrule
        Base Model
        & 0.023 \\
        Base + Intent Token
        & 0.029 \\
        Base + Intent and Policy Tokens
        & 0.032 \\
        Direct LLM Inference
        & 4.437 \\
        \bottomrule
    \end{tabular}
\end{table}

\section{Conclusion}


In this paper, we identify the \emph{Understanding--Action Gap} in generative recommendation and distinguish task-oriented intent from recommendation policy. We propose a feedback-grounded framework that induces candidate policies from positive behavioral transitions, evaluates their incremental utility over an intent-only baseline, and iteratively refines them through group-wise execution feedback. This process enables the model to retain decision principles with verified recommendation gains rather than relying solely on linguistically plausible LLM outputs. To support efficient deployment, we transfer intent and validated policy knowledge into an ordered latent chain within a lightweight SID recommender through dual-space relational distillation, enabling LLM-free online inference. Experiments on public benchmarks demonstrate consistent improvements across datasets and backbones, while large-scale online A/B tests achieve gains of 4.506\% in Revenue and 4.621\% in ADVV. These results demonstrate that validated policy knowledge complements intent understanding and enables accurate, efficient, and LLM-free generative recommendation.


%
\bibliographystyle{ACM-Reference-Format}
\bibliography{main}

\appendix

\section{Additional Experimental Details}
\label{app:experimental_details}

\subsection{Dataset Statistics}
\label{app:dataset_details}

The Amazon Review datasets contain user--item interactions and item metadata, and have been extensively used in sequential and SID-based
generative recommendation~\cite{tiger,wang2024letter}. For each user, we sort
the interactions chronologically. The last interaction is held out for
testing, the second-to-last interaction is used for validation, and all
preceding interactions are used for training. We apply the same preprocessing
protocol and data splits to all compared methods.
Table~\ref{tab:dataset_statistics} reports the statistics of the processed
datasets.


\subsection{Implementation Details}
\label{app:implementation_details}

Our method is implemented in PyTorch and trained on eight NVIDIA A800 GPUs,
each with 80\,GB of memory. The learning rate is selected from
$\{1\times10^{-3},\,2\times10^{-3}\}$, the batch size is set to 2048, and
the number of training epochs is selected from $\{200,300\}$. The intent- and
policy-distillation weights, $\lambda_I$ and $\lambda_P$, are independently
selected from $\{0.5,1.0\}$. The Policy Agent generates $K=5$ candidate
policies in each discovery round. Although we analyze additional rounds to
study the policy-evolution trend, we use at most $R=2$ refinement rounds in
practice to balance recommendation quality and computational cost. For
relational distillation, the numbers of first- and higher-order neighbors are
set to $K_1=K_2=5$, and the higher-order coefficient is fixed to
$\gamma=0.1$. All experiments are repeated five times, and the average results
are reported.

For traditional recommendation baselines, we follow exactly the same
experimental settings as prior studies when they use the same datasets, data
splits, and evaluation metrics~\cite{tiger,hou2025generating,hou2025actionpiece}.
TIGER~\cite{tiger} follows the implementation and training configuration
reported by GRID~\cite{ju2025handbook}, while
LETTER~\cite{wang2024letter} is reproduced using its official code and
recommended hyperparameters. For fair comparison, all methods use the same preprocessed interaction sequences and data splits.

\subsection{Baseline Details}
\label{app:baseline_details}

We consider conventional recommendation, SID-based generative recommendation, and direct LLM inference
baselines.

\begin{itemize}
    \item \textbf{Caser}~\cite{tang2018caser} embeds recent interactions as
    a two-dimensional sequence representation and applies convolutional filters
    to capture both sequential patterns and user preferences.

    \item \textbf{GRU4Rec}~\cite{hidasi2016gru4rec} applies gated recurrent
    units to model interaction sequences for session-based next-item
    recommendation.

    \item \textbf{SASRec}~\cite{kang2018self} uses causal self-attention to
    identify the historical interactions most relevant to predicting the next
    item.

    \item \textbf{BERT4Rec}~\cite{sun2019bert4rec} learns bidirectional
    sequence representations through a Cloze-style masked-item prediction
    objective.

    \item \textbf{HGN}~\cite{ma2019hgn} combines feature-level and
    instance-level gating to model long- and short-term user interests, while
    explicitly capturing relations between historical and candidate items.

    \item \textbf{P5}~\cite{geng2022recommendation} formulates multiple
    recommendation tasks within a unified text-to-text framework using
    personalized natural-language prompts.

    \item \textbf{TIGER}~\cite{tiger} represents each item as a hierarchical
    Semantic ID and trains a sequence-to-sequence model to autoregressively
    generate the Semantic ID of the next item from the user's interaction
    history.

    \item \textbf{LETTER}~\cite{wang2024letter} learns an item tokenizer that
    jointly incorporates hierarchical semantics, collaborative signals, and
    code-assignment diversity for generative recommendation.

    \item \textbf{Direct LLM} uses
    Qwen3.5-122B-A10B~\cite{yang2025qwen3} to directly generate a textual profile of
    the next item from the interaction history. We retrieve recommendations by
    matching the generated profile against candidate-item representations
    using embedding similarity.
\end{itemize}

\begin{table}[t]
    \centering
    \caption{Statistics of the datasets.}
    \label{tab:dataset_statistics}
    \begin{tabular}{lcccc}
        \toprule
        \textbf{Dataset} & \textbf{\#Users} & \textbf{\#Items} & \textbf{\#Interactions} & \textbf{Avg. Length} \\
        \midrule
        Beauty & 22,363 & 12,101 & 176,139 & 8.87 \\
        Toys   & 19,412 & 11,924 & 148,185 & 8.63 \\
        Sports & 35,598 & 18,357 & 260,739 & 8.32 \\
        \bottomrule
    \end{tabular}
\end{table}

\begin{table}[t]
    \centering
    \caption{
    Ablation of policy selection and evolution on Toys with TIGER as the backbone.
    }
    \label{tab:rq3_evolution_ablation_toys}
    \begin{tabular}{lcccc}
        \toprule
        \textbf{Policy Construction}
        & \textbf{R@5}
        & \textbf{R@10}
        & \textbf{N@5}
        & \textbf{N@10} \\
        \midrule
        \textbf{Full Model}
        & \textbf{0.0485}
        & \textbf{0.0712}
        & \textbf{0.0322}
        & \textbf{0.0393} \\
        w/o Policy Evolution
        & 0.0471
        & 0.0702
        & 0.0319
        & 0.0390 \\
        Random Policy Selection
        & 0.0468
        & 0.0666
        & 0.0308
        & 0.0379 \\
        Target-as-Policy
        & 0.0466
        & 0.0696
        & 0.0310
        & 0.0381 \\
        \bottomrule
    \end{tabular}
\end{table}

\begin{table}[t]
    \centering
    \small
    \setlength{\tabcolsep}{3pt}
    \caption{
    Effect of policy token and policy distillation on Toys with TIGER. All variants retain the Intent Token and intent distillation.
    }
    \label{tab:rq2_component_ablation_toys}
    \begin{tabular}{cc|cccc}
        \toprule
        \textbf{Policy Token}
        & \textbf{Policy Distillation}
        & \textbf{R@5}
        & \textbf{R@10}
        & \textbf{N@5}
        & \textbf{N@10} \\
        \midrule
        -- & --
        & 0.0444
        & 0.0668
        & 0.0296
        & 0.0362 \\
        \checkmark & --
        & 0.0453
        & 0.0688
        & 0.0309
        & 0.0379 \\
        \checkmark & \checkmark
        & \textbf{0.0485}
        & \textbf{0.0712}
        & \textbf{0.0322}
        & \textbf{0.0393} \\
        \bottomrule
    \end{tabular}
\end{table}

\section{Additional Experimental Results}
\label{app:additional_results}

This section provides additional experimental results complementing the main
analysis. We first verify the roles of intent, policy, and feedback-grounded
policy discovery on Toys and Games and Sports and Outdoors datasets. We then report
additional ablations of relational distillation and visualize the relational
structure learned in the Intent Token space.

\subsection{Results on Additional Datasets}
\label{app:additional_dataset_results}

Tables~\ref{tab:rq3_evolution_ablation_toys}-\ref{tab:distillation_sports} report additional results on Toys and Games and
Sports and Outdoors. The trends are consistent with those on Beauty: explicit
policy supervision outperforms merely introducing an additional latent token,
outcome-grounded policy selection and evolution yield stronger policies than
the alternative construction strategies, and relational distillation
consistently surpasses direct representation matching.

\begin{table}[t]
    \centering
    \caption{
    Ablation study of knowledge distillation on Toys with TIGER.
    }
    \label{tab:distillation_toys}
    \setlength{\tabcolsep}{2.5pt}
    \begin{tabular}{lcccc}
        \toprule
        \textbf{Method}
        & \textbf{R@5}
        & \textbf{R@10}
        & \textbf{N@5}
        & \textbf{N@10} \\
        \midrule
        \textbf{Relational Distillation (Ours)}
        & \textbf{0.0485}
        & \textbf{0.0712}
        & \textbf{0.0322}
        & \textbf{0.0393} \\
        Direct Distillation
        & 0.0473
        & 0.0684
        & 0.0308
        & 0.0381 \\
        w/o Higher-Order Distillation & 0.0478 & 0.0687 & 0.0319 & 0.0387 \\
        w/o Intent Distillation
        & 0.0460
        & 0.0690
        & 0.0313
        & 0.0386 \\
        w/o Policy Distillation
        & 0.0453
        & 0.0688
        & 0.0309
        & 0.0379 \\
        w/o Distillation
        & 0.0430
        & 0.0662
        & 0.0289
        & 0.0363 \\
        \bottomrule
    \end{tabular}
\end{table}

\begin{table}[t]
    \centering
    \caption{
    Ablation of policy selection and evolution on Sports with TIGER as the backbone.
    }
    \label{tab:rq3_evolution_ablation_sports}
    \begin{tabular}{lcccc}
        \toprule
        \textbf{Policy Construction}
        & \textbf{R@5}
        & \textbf{R@10}
        & \textbf{N@5}
        & \textbf{N@10} \\
        \midrule
        \textbf{Full Model}
        & \textbf{0.0288}
        & \textbf{0.0441}
        & \textbf{0.0186}
        & \textbf{0.0234} \\
        w/o Policy Evolution
        & 0.0280
        & 0.0433
        & 0.0183
        & 0.0233 \\
        Random Policy Selection
        & 0.0272
        & 0.0426
        & 0.0178
        & 0.0227 \\
        Target-as-Policy
        & 0.0279
        & 0.0424
        & 0.0181
        & 0.0228 \\
        \bottomrule
    \end{tabular}
\end{table}

\begin{table}[t]
    \centering
    \small
    \setlength{\tabcolsep}{3pt}
    \caption{
    Effect of policy token and policy distillation on Sports with TIGER. All variants retain the Intent Token and intent distillation.
    }
    \label{tab:rq2_component_ablation_sports}
    \begin{tabular}{cc|cccc}
        \toprule
        \textbf{Policy Token}
        & \textbf{Policy Distillation}
        & \textbf{R@5}
        & \textbf{R@10}
        & \textbf{N@5}
        & \textbf{N@10} \\
        \midrule
        -- & --
        & 0.0266
        & 0.0416
        & 0.0176
        & 0.0223 \\
        \checkmark & --
        & 0.0283
        & 0.0430
        & 0.0185
        & 0.0233 \\
        \checkmark & \checkmark
        & \textbf{0.0288}
        & \textbf{0.0441}
        & \textbf{0.0186}
        & \textbf{0.0234} \\
        \bottomrule
    \end{tabular}
\end{table}

\begin{table}[t]
    \centering
    \caption{
    Ablation study of knowledge distillation on Sports with TIGER.
    }
    \label{tab:distillation_sports}
    \setlength{\tabcolsep}{2.5pt}
    \begin{tabular}{lcccc}
        \toprule
        \textbf{Method}
        & \textbf{R@5}
        & \textbf{R@10}
        & \textbf{N@5}
        & \textbf{N@10} \\
        \midrule
        \textbf{Relational Distillation (Ours)}
        & \textbf{0.0288}
        & \textbf{0.0441}
        & \textbf{0.0186}
        & \textbf{0.0234} \\
        Direct Distillation
        & 0.0274
        & 0.0434
        & 0.0180
        & 0.0231 \\
        w/o Higher-Order Distillation & 0.0279 & 0.0424 & 0.0183 & 0.0228 \\
        w/o Intent Distillation
        & 0.0277
        & 0.0421
        & 0.0184
        & 0.0230 \\
        w/o Policy Distillation
        & 0.0283
        & 0.0430
        & 0.0185
        & 0.0233 \\
        w/o Distillation
        & 0.0268
        & 0.0416
        & 0.0174
        & 0.0222 \\
        \bottomrule
    \end{tabular}
\end{table}

\subsection{Intent-Space Structure Visualization}
\label{app:intent_structure_visualization}

\begin{figure}[t]
    \centering
    \includegraphics[width=\linewidth]{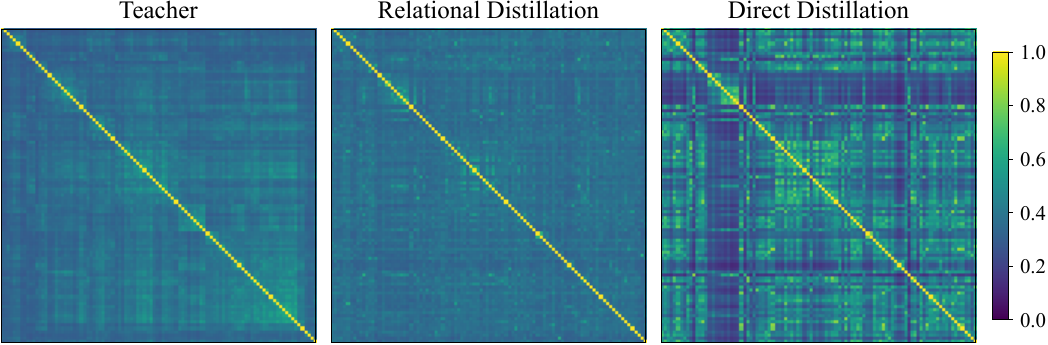}
    \caption{
   Pairwise similarity structures. From left to right: the teacher intent space, our relationally distilled Intent Token space, and the directly distilled Intent Token space.
    }
    \label{fig:intent_structure}
\end{figure}

Figure~\ref{fig:intent_structure} visualizes the pairwise similarity structures
in the intent space. Relational distillation more faithfully preserves the
teacher-induced user structure than direct point-wise distillation, consistent
with the policy-space visualization reported in the main text.

\section{Case Study}

\subsection{Case Study: Complementary Roles of Intent and Policy}

We present a representative case from the Beauty dataset to illustrate the distinct yet complementary roles of intent and policy. The intent identifies the user's current demand and narrows the recommendation space to a semantically relevant region. However, multiple candidate actions may remain plausible within this region. The policy further determines which behavioral transition should be prioritized and which intent-consistent but contextually inappropriate alternatives should be suppressed.

\paragraph{Completing a Gel-Nail Workflow.}
The user previously interacted with a nail file, nail-dotting tools, and gel nail polish, while the ground-truth next item is a one-hand LED gel-nail dryer. Based on the interaction history, the intent correctly identifies nail care and creative nail art as the user's current demand, thereby restricting the candidate space to nail-related products. Nevertheless, this semantic region still contains several plausible recommendations, including additional nail polish, decorative tools, brush sets, and curing equipment.

Conditioned only on the intent, the model recommends a nail-art brush set and obtains a semantic reward of 0.5941. Although this prediction is consistent with the general nail-art demand, it overlooks the immediate functional dependency introduced by the user's recent interaction with gel nail polish. In contrast, the policy identifies curing as the next required step in the gel-nail workflow and suppresses products that merely repeat functions already covered by the user's history. Guided by this policy, the model recommends the ground-truth LED gel-nail dryer, increasing the semantic reward to 0.8754.

\begin{table}[t]
\centering
\caption{The inferred intent and outcome-validated policy for a representative gel-nail workflow case.}
\small
\begin{tabular}{p{0.23\linewidth} p{0.69\linewidth}}
\toprule
\textbf{Component} & \textbf{Content} \\
\midrule
Intent &
The user is currently completing a gel-nail care and creative nail-art workflow. \\

Recommendation Direction &
Prioritize products that provide a direct functional complement to the most recently interacted item, particularly tools required for the immediate next step of the gel-nail workflow, such as curing equipment following gel-polish application. \\

Rejection Boundary &
Avoid products that are only broadly related to nail art but do not advance the current workflow, as well as tools that duplicate functions already covered by the user's previous interactions. \\
\bottomrule
\end{tabular}
\label{tab:intent_policy_case}
\end{table}

This case demonstrates that the distinction between intent and policy is not merely one of semantic granularity. Intent summarizes the user's current demand state and identifies a plausible candidate region. Policy instead acts as a history-dependent decision rule within that region: it specifies which functional transition should be prioritized and which semantically relevant but action-inappropriate alternatives should be rejected. In this example, recognizing that the user is engaged in gel nail art does not uniquely determine the next recommendation. The policy supplies the missing action-level guidance by identifying curing as the immediate functional step following gel-polish application.

\subsection{Case Study: Feedback-Driven Policy Evolution}

\label{app:policy_evolution_case}

We present a representative example from the Beauty dataset to illustrate one complete round of feedback-driven policy evolution. The user previously interacted with wrinkle-treatment patches and a bentonite-clay facial cleanser. The ground-truth next item is a rose-water facial hydrator. Importantly, this item is used only by the external reward environment and is never included in the prompts of either the Policy Agent or the Feedback Agent.
\begin{table}[t]
    \centering
    \small
    \setlength{\tabcolsep}{5pt}
    \caption{Inferred intent and intent-only prediction for the policy-evolution case.}
    \label{tab:evolution_case_intent}
    \begin{tabular}{p{0.30\linewidth}p{0.60\linewidth}}
        \toprule
        \textbf{Component} & \textbf{Content} \\
        \midrule
        Intent
        &
        The user seeks natural, ingredient-focused skincare solutions for
        addressing facial wrinkles and deep cleansing. \\
        \midrule
        Intent-only Prediction
        &
        Bentonite-clay face mask. \\
        \bottomrule
    \end{tabular}
\end{table}
The intent correctly localizes the facial-skincare demand, but the prediction remains dominated by the most recent deep-cleansing behavior.

\paragraph{Initial policy hypotheses.}
Conditioned on the same history and intent, the Policy Agent generates exactly $K=5$ behaviorally distinct candidate policies. The original outputs are slightly shortened for readability without altering their decision semantics.

\begin{table}[t]
    \centering
    \small
    \setlength{\tabcolsep}{5pt}
    \caption{
    Five behaviorally distinct policy hypotheses generated for the same user state. The outputs are lightly shortened for readability without changing their decision semantics.
    }
    \label{tab:case_initial_policies}
    \begin{tabular}{p{0.45\textwidth}}
        \toprule
        \textbf{Policy Hypothesis} \\
        \midrule

        Prioritize natural ingredient-based deep-cleansing masks or clay
        products for the full face. Avoid synthetic exfoliants or products
        inconsistent with the natural, earth-derived preference. \\

        \midrule

        Prioritize cleansing or detoxifying products that prepare the skin for
        anti-aging treatment. Avoid products offering moisturization without
        active cleansing or detoxifying effects. \\

        \midrule

        Prioritize face cleansers or masks that differ from the previous
        eye-specific treatment. Avoid additional eye patches or products that
        duplicate its localized function. \\

        \midrule

        Prioritize natural products for physical exfoliation, pore cleansing,
        or texture smoothing. Avoid hydration-only products without texture
        benefits. \\

        \midrule

        Prioritize complementary steps in a skincare routine, such as a
        deep-cleansing mask following a targeted treatment. Avoid products that
        do not advance the sequential routine. \\

        \bottomrule
    \end{tabular}
\end{table}
The five policies lead to heterogeneous execution outcomes. Policies
emphasizing routine progression and functional complementarity improve over
the intent-only branch, whereas rigid constraints on ingredient origin,
hydration, or physical exfoliation are detrimental. The routine-completion policy is selected as the best initial policy. Nevertheless, it still predicts
another clay mask, revealing that the initial hypothesis set remains overly
concentrated on masks and deep-cleansing formats.

\paragraph{Group-wise feedback.}
The Feedback Agent jointly receives the five
policy--prediction--advantage triplets. It does not observe the last item in the user's training sequence,
its metadata, or its semantic representation. By comparing successful and
unsuccessful candidates, it produces the following structured critique.
For readability, repeated entries within each original feedback field are
merged without changing their decision semantics.

\begin{table}[t]
    \centering
    \small
    \setlength{\tabcolsep}{5pt}
    \caption{
    Structured group-wise critique derived from the five initial policies.
    }
    \label{tab:case_group_critique}
    \begin{tabular}{p{0.25\linewidth}p{0.65\linewidth}}
        \toprule
        \textbf{Component} & \textbf{Content} \\
        \midrule
        Preserve Patterns
        &
        Preserve functional complementarity and progression from a localized
        treatment toward the next full-face skincare step, rather than simply
        repeating the latest product category. \\
        \midrule
        Shared Failure Modes
        &
        Avoid over-constraining recommendations by ingredient origin or
        physical texture, which narrows the decision space and weakens
        functional alignment with the current routine. \\
        \midrule
        Underexplored Directions
        &
        Explore non-mask formats, including serums, toners, and cleansers, that
        can bridge the previous targeted treatment and subsequent full-face
        care. \\
        \midrule
        Regeneration Guidance
        &
        Relax natural-versus-synthetic constraints, expand beyond masks and
        scrubs, and prioritize the sequential function of the next product over
        surface attributes such as texture or format. \\
        \bottomrule
    \end{tabular}
\end{table}

The critique does not identify or recommend a specific item. Instead, it identifies which decision principles distinguish beneficial policies from semantically
plausible but ineffective ones: functional progression should be preserved,
whereas repeated formats and unnecessary surface-level constraints should be
corrected.

\paragraph{First-round evolution.}
Conditioned on the group-wise critique, the Policy Agent generates a new set of
candidate policies. All candidates are executed under the same controlled
setting, and the candidate with the highest verified advantage is selected.
For conciseness, we report the selected first-round policy below.

\begin{table}[t]
    \centering
    \small
    \setlength{\tabcolsep}{5pt}
    \caption{Selected policy after the first feedback-driven evolution round.}
    \label{tab:evolution_case_refined_policy}
    \begin{tabular}{p{0.30\linewidth}p{0.60\linewidth}}
        \toprule
        \textbf{Component} & \textbf{Content} \\
        \midrule
        Recommendation Direction
        &
        Prioritize functional progression through non-mask formats, such as
        toners, serums, or gentle cleansers, that bridge the targeted eye
        treatment and subsequent full-face care. \\
        \midrule
        Rejection Boundary
        &
        Avoid additional eye-specific products that duplicate the same
        localized function, and do not exclude effective products solely
        because they lack natural branding. \\
        \bottomrule
    \end{tabular}
\end{table}

The evolved policy predicts a leave-on facial serum rather than another clay
mask. Its reward increases from 0.5854 for the best initial policy to 0.6765, exceeding the acceptance criterion and therefore passing the outcome-based gate. The prediction is not an exact match to the observed
toner; instead, the evolution moves the recommendation from a repetitive mask
toward a more appropriate non-mask, full-face treatment region.

Overall, this case illustrates the complete evolution process. Heterogeneous
policy executions provide comparative evidence, the Feedback Agent summarizes
their shared strengths and failure modes, and the Policy Agent converts this
feedback into a revised recommendation direction and rejection boundary. The
revision is retained only after its executed consequence is verified.

\section{Prompt Templates}
\label{app:prompts}

This section presents the prompts used by the Intent Agent, Policy Agent,
controlled executor, and Feedback Agent.

\subsection{Task-Oriented Intent Induction}
\label{app:prompt_intent}

This prompt infers the task-oriented demand most likely to drive the user's
next interaction.

\begin{tcolorbox}[
    breakable,
    colback=gray!10,
    colframe=black,
    boxrule=0.8pt,
    arc=2pt,
    left=4pt,
    right=4pt,
    top=4pt,
    bottom=4pt
]
\textbf{System Prompt:}
You are an Intent Agent for next-item recommendation in [DOMAIN].

Given a chronological interaction history and the metadata of its items, infer
the task-oriented intent most likely to drive the user's immediate next
interaction.

The intent should describe the user's current demand rather than a broad
long-term profile. Infer it only from the supplied history and metadata.
Describe what the user is currently trying to achieve, complete, maintain,
replace, continue, or explore. Do not predict a concrete item, title, category
path, item ID, or ASIN. Do not introduce unsupported demographic, medical,
compatibility, or lifestyle assumptions. Output one concise intent in valid
JSON.

\medskip

\textbf{User Content:}

\begin{Verbatim}[fontsize=\scriptsize]
<User_History_Sequence>
[CHRONOLOGICAL ITEM POSITIONS]
</User_History_Sequence>

<Historical_Item_Information>
[TITLE AND CATEGORY OF EACH HISTORICAL ITEM]
</Historical_Item_Information>

Infer the task-oriented demand most likely to drive the immediate next
interaction.

Return exactly:
{
  "intent": "..."
}
\end{Verbatim}
\end{tcolorbox}

\subsection{Initial Policy Hypothesis Generation}
\label{app:prompt_policy_initialization}

This prompt generates \(K\) behaviorally distinct policy hypotheses
conditioned on the user history and inferred Intent.

\begin{tcolorbox}[
    breakable,
    colback=gray!10,
    colframe=black,
    boxrule=0.8pt,
    arc=2pt,
    left=4pt,
    right=4pt,
    top=4pt,
    bottom=4pt
]
\textbf{System Prompt:}
You are a Policy Agent for personalized next-item recommendation in [DOMAIN].

A policy is an executable decision rule consisting of a recommendation
direction and a rejection boundary. The recommendation direction specifies
which history-to-next-item transition, functional relation, or task step to
prioritize. The rejection boundary specifies which plausible but unsuitable
alternatives to suppress.

Given a user history and an inferred intent, generate exactly five behaviorally
distinct Policy hypotheses. Each policy must describe how the recommender
should act rather than restating the Intent. Ground every Policy in the supplied
History and Intent, include both what to prioritize and what to avoid, and keep
the rejection boundary local to the current decision. Do not predict or
describe a hidden target, unseen item, item ID, or unsupported attribute.
Output valid JSON only.

\medskip

\textbf{User Content:}

\begin{Verbatim}[fontsize=\scriptsize]
<User_History_Sequence>
[CHRONOLOGICAL ITEM POSITIONS]
</User_History_Sequence>

<Historical_Item_Information>
[TITLE AND CATEGORY OF EACH HISTORICAL ITEM]
</Historical_Item_Information>

<Inferred_User_Intent>
[INTENT]
</Inferred_User_Intent>

Generate exactly five behaviorally distinct Policy hypotheses.

Return exactly:
{
  "policies": [
    {
      "policy_id": "p1",
      "recommendation_direction": "...",
      "rejection_boundary": "...",
      "policy_value": "Prioritize ... Avoid ..."
    },
    {
      "policy_id": "p2",
      "recommendation_direction": "...",
      "rejection_boundary": "...",
      "policy_value": "Prioritize ... Avoid ..."
    },
    {
      "policy_id": "p3",
      "recommendation_direction": "...",
      "rejection_boundary": "...",
      "policy_value": "Prioritize ... Avoid ..."
    },
    {
      "policy_id": "p4",
      "recommendation_direction": "...",
      "rejection_boundary": "...",
      "policy_value": "Prioritize ... Avoid ..."
    },
    {
      "policy_id": "p5",
      "recommendation_direction": "...",
      "rejection_boundary": "...",
      "policy_value": "Prioritize ... Avoid ..."
    }
  ]
}
\end{Verbatim}
\end{tcolorbox}

\subsection{Controlled Policy Execution}
\label{app:prompt_policy_execution}

This prompt predicts one semantic next-item profile in each independent model
call. We run it once without a policy for the intent-only baseline and once for
each candidate policy.

\begin{tcolorbox}[
    breakable,
    colback=gray!10,
    colframe=black,
    boxrule=0.8pt,
    arc=2pt,
    left=4pt,
    right=4pt,
    top=4pt,
    bottom=4pt
]
\textbf{System Prompt:}
You are a controlled next-item prediction executor for [DOMAIN].

Given a user History, an inferred Intent, and an optional recommendation
Policy, predict exactly one semantic profile describing the user’s immediate next item. When no Policy is provided, make the prediction using only the History
and Intent. When a policy is provided, use it as an additional decision rule.

The Policy should guide selection but must not override explicit behavioral
evidence. Do not copy a historical item title. Do not introduce unsupported
brands, ingredients, functions, or attributes. Do not output rationales,
confidence scores, target guesses, item IDs, or ASINs. Output valid JSON only.

\medskip

\textbf{User Content for the Intent-Only Branch:}

\begin{Verbatim}[fontsize=\scriptsize]
<User_History_Sequence>
[CHRONOLOGICAL ITEM POSITIONS]
</User_History_Sequence>

<Historical_Item_Information>
[TITLE AND CATEGORY OF EACH HISTORICAL ITEM]
</Historical_Item_Information>

<Inferred_User_Intent>
[INTENT]
</Inferred_User_Intent>

No recommendation Policy is provided.

Return exactly:
{
  "predicted_title": "...",
  "predicted_category": "..."
}
\end{Verbatim}

\medskip

\textbf{User Content for a Policy-Conditioned Branch:}

\begin{Verbatim}[fontsize=\scriptsize]
<User_History_Sequence>
[CHRONOLOGICAL ITEM POSITIONS]
</User_History_Sequence>

<Historical_Item_Information>
[TITLE AND CATEGORY OF EACH HISTORICAL ITEM]
</Historical_Item_Information>

<Inferred_User_Intent>
[INTENT]
</Inferred_User_Intent>

<Candidate_Policy>
Policy ID: [POLICY ID]
Policy: [POLICY TEXT]
</Candidate_Policy>

Return exactly:
{
  "policy_id": "[POLICY ID]",
  "predicted_title": "...",
  "predicted_category": "..."
}
\end{Verbatim}
\end{tcolorbox}

\subsection{Group-Wise Policy Feedback}
\label{app:prompt_policy_feedback}

This prompt compares the independently executed policy candidates and produces
one target-blind group-level Critique.

\begin{tcolorbox}[
    breakable,
    colback=gray!10,
    colframe=black,
    boxrule=0.8pt,
    arc=2pt,
    left=4pt,
    right=4pt,
    top=4pt,
    bottom=4pt
]
\textbf{System Prompt:}
You are a target-blind Feedback Agent for Policy discovery in [DOMAIN].

You receive several candidate Policies evaluated independently under the same
hidden user state. For each candidate, you may observe only its Policy text,
executed semantic prediction, and scalar advantage over the Intent-only
baseline. You never observe the user History, inferred Intent, target item,
target metadata, or target representation.

Compare all candidates jointly. Identify decision patterns associated with
stronger outcomes, shared failure modes among weaker candidates, overly
restrictive rejection boundaries, and useful directions that remain insufficiently explored by the group. Ground conclusions in candidate IDs and relative
advantage patterns.

Do not copy concrete predicted titles, brands, full category paths, or uniquely
identifying attribute combinations. Do not infer or reconstruct the hidden
target, and do not prescribe a concrete next item. Output valid JSON only.

\medskip

\textbf{User Content:}

\begin{Verbatim}[fontsize=\scriptsize]
<Evaluated_Policy_Group>
Candidate ID: [ID 1]
Policy: [POLICY 1]
Executed prediction: [PREDICTION 1]
Scalar advantage: [ADVANTAGE 1]

Candidate ID: [ID 2]
Policy: [POLICY 2]
Executed prediction: [PREDICTION 2]
Scalar advantage: [ADVANTAGE 2]

...

Candidate ID: [ID 5]
Policy: [POLICY 5]
Executed prediction: [PREDICTION 5]
Scalar advantage: [ADVANTAGE 5]
</Evaluated_Policy_Group>

Return exactly:
{
  "preserve_patterns": ["..."],
  "shared_failure_modes": ["..."],
  "underexplored_directions": ["..."],
  "regeneration_guidance": ["..."],
  "comparative_evidence": [
    {
      "candidate_ids": ["..."],
      "advantage_pattern": "...",
      "policy_level_inference": "..."
    }
  ]
}
\end{Verbatim}
\end{tcolorbox}


\subsection{Feedback-Guided Policy Refinement}
\label{app:prompt_policy_refinement}

This prompt generates a new set of \(K=5\) policies conditioned on the
group-level Critique.

\begin{tcolorbox}[
    breakable,
    colback=gray!10,
    colframe=black,
    boxrule=0.8pt,
    arc=2pt,
    left=4pt,
    right=4pt,
    top=4pt,
    bottom=4pt
]
\textbf{System Prompt:}
You are a Policy Agent for feedback-guided Policy refinement in [DOMAIN].

Given a user History, inferred Intent, current Policy hypotheses, and a
target-blind group Critique, generate exactly five new and behaviorally
distinct Policy hypotheses.

Preserve decision patterns supported by the Critique, correct the diagnosed
failure modes, and explore useful but underrepresented directions. Use the
Critique only as meta-level guidance and ground every new Policy in the supplied
History and Intent. Each Policy must contain a recommendation direction, a
rejection boundary, and a concise policy value beginning with "Prioritize" and
containing an "Avoid" clause.

Do not restate the Intent, copy previous predictions, infer the hidden target,
or mention unseen item titles, item IDs, ASINs, or unsupported attributes.
Keep rejection boundaries local to the current decision. Output valid JSON
only.

\medskip

\textbf{User Content:}

\begin{Verbatim}[fontsize=\scriptsize]
<User_History_Sequence>
[CHRONOLOGICAL ITEM POSITIONS]
</User_History_Sequence>

<Historical_Item_Information>
[TITLE AND CATEGORY OF EACH HISTORICAL ITEM]
</Historical_Item_Information>

<Inferred_User_Intent>
[INTENT]
</Inferred_User_Intent>

<Current_Policy_Group>
[CURRENT POLICY GROUP]
</Current_Policy_Group>

<Target_Blind_Group_Critique>
[GROUP CRITIQUE JSON]
</Target_Blind_Group_Critique>

Generate exactly five new and behaviorally distinct Policy hypotheses.

Return exactly:
{
  "policies": [
    {
      "policy_id": "new_1",
      "recommendation_direction": "...",
      "rejection_boundary": "...",
      "policy_value": "Prioritize ... Avoid ..."
    },
    {
      "policy_id": "new_2",
      "recommendation_direction": "...",
      "rejection_boundary": "...",
      "policy_value": "Prioritize ... Avoid ..."
    },
    {
      "policy_id": "new_3",
      "recommendation_direction": "...",
      "rejection_boundary": "...",
      "policy_value": "Prioritize ... Avoid ..."
    },
    {
      "policy_id": "new_4",
      "recommendation_direction": "...",
      "rejection_boundary": "...",
      "policy_value": "Prioritize ... Avoid ..."
    },
    {
      "policy_id": "new_5",
      "recommendation_direction": "...",
      "rejection_boundary": "...",
      "policy_value": "Prioritize ... Avoid ..."
    }
  ]
}
\end{Verbatim}
\end{tcolorbox}

\end{document}